\documentclass[10pt]{article}

\usepackage{color,amsmath,latexsym,amssymb,revsymb}
\usepackage{theorem}

\newtheorem{definition}{Definition}
\newtheorem{proposition}[definition]{Proposition}
\newtheorem{lemma}[definition]{Lemma}

\newtheorem{theorem}[definition]{Theorem}
\newtheorem{corollary}[definition]{Corollary}

\newtheorem{remark}[definition]{Remark}
\newtheorem{example}[definition]{Example}

\def\squareforqed{\hbox{\rlap{$\sqcap$}$\sqcup$}}
\def\qed{\ifmmode\squareforqed\else{\unskip\nobreak\hfil
\penalty50\hskip1em\null\nobreak\hfil\squareforqed
\parfillskip=0pt\finalhyphendemerits=0\endgraf}\fi}
\def\endenv{\ifmmode\;\else{\unskip\nobreak\hfil
\penalty50\hskip1em\null\nobreak\hfil\;
\parfillskip=0pt\finalhyphendemerits=0\endgraf}\fi}

\mathchardef\ordinarycolon\mathcode`\: \mathcode`\:=\string"8000
\def\vcentcolon{\mathrel{\mathop\ordinarycolon}}
\begingroup \catcode`\:=\active
  \lowercase{\endgroup
  \let :\vcentcolon
  }

\newcommand{\nc}{\newcommand}
\nc{\rnc}{\renewcommand} \nc{\beq}{\begin{equation}}
\nc{\eeq}{{\end{equation}}} \nc{\beqa}{\begin{eqnarray}}
\nc{\eeqa}{\end{eqnarray}} \nc{\lbar}[1]{\overline{#1}}
\nc{\bra}[1]{\langle#1|} \nc{\ket}[1]{|#1\rangle}
\nc{\ketbra}[2]{|#1\rangle\!\langle#2|}
\nc{\braket}[2]{\langle#1|#2\rangle} \nc{\proj}[1]{|
#1\rangle\!\langle #1 |} \nc{\avg}[1]{\langle#1\rangle}
\rnc{\max}{\operatorname{max}} \nc{\Rank}{\operatorname{Rank}}
\nc{\smfrac}[2]{\mbox{$\frac{#1}{#2}$}} \nc{\tr}{\operatorname{Tr}}
\nc{\ox}{\otimes} \nc{\dg}{\dagger} \nc{\dn}{\downarrow}
\nc{\cA}{{\cal A}} \nc{\cB}{{\cal B}} \nc{\cC}{{\cal C}}
\nc{\cD}{{\cal D}} \nc{\cE}{{\cal E}} \nc{\cF}{{\cal F}}
\nc{\cG}{{\cal G}} \nc{\cH}{{\cal H}} \nc{\cI}{{\cal I}}
\nc{\cJ}{{\cal J}} \nc{\cK}{{\cal K}} \nc{\cL}{{\cal L}}
\nc{\cM}{{\cal M}} \nc{\cN}{{\cal N}} \nc{\cO}{{\cal O}}
\nc{\cP}{{\cal P}} \nc{\cR}{{\cal R}} \nc{\cS}{{\cal S}}
\nc{\cT}{{\cal T}} \nc{\cX}{{\cal X}} \nc{\cZ}{{\cal Z}}
\nc{\csupp}{{\operatorname{csupp}}}
\nc{\qsupp}{{\operatorname{qsupp}}} \nc{\var}{\operatorname{var}}
\nc{\rar}{\rightarrow} \nc{\lrar}{\longrightarrow}
\nc{\polylog}{\operatorname{polylog}} \nc{\1}{{\openone}}

\def\a{\alpha}
\def\b{\beta}
\def\g{\gamma}

\def\ve{\varepsilon}

\def\r{\rho}

\def\ph{\varphi}

\def\ps{\psi}

\def\Ph{\Phi}
\def\Ps{\Psi}

\newcommand{\lin}{{\operatorname{span}\,}}

\usepackage{bm}
\bmdefine{\bbb}{b}
\bmdefine{\uuu}{u} \bmdefine{\vvv}{v} \bmdefine{\www}{w}
\bmdefine{\eee}{e} \bmdefine{\xxx}{x} \bmdefine{\zerovec}{0}

\usepackage{qite}
\begin{document}

\title{
  Tensor rank problem in statistical high-dimensional data
  and quantum information theory:their comparisons on the methods and the results
}
\author{
  Toshio Sakata\makebox[0pt][l]{\addressmark{1}} \\
  sakata@design.kyushu-u.ac.jp
  \and
  Lin Chen\makebox[0pt][l]{\addressmark{2}}\\
  cqtcl@nus.edu.sg
  \and
   Toshio Sumi\makebox[0pt][l]{\addressmark{1}}\\
  sumi@design.kyushu-u.ac.jp
  \and
     Mitsuhiro Miyazaki\makebox[0pt][l]{\addressmark{3}}\\
 g53448@kyokyo-u.ac.jp
}

\seteaddress{
\makebox[0pt][r]{\addressmark{1}}%
Dept. of  Human Science, Faculty of Design,  Kyushu University \\
4-9-1 Shiobaru Minami-ku Fukuoka, Japan\\
TEL: +81-92-651-4450 \quad FAX: \\
\makebox[0pt][r]{\addressmark{2}}%
Center for Quantum Technologies,  National University of Singapore \\
3 Science Drive 2, Singapore\\
TEL:65-65164476  \quad FAX: \\
\makebox[0pt][r]{\addressmark{3}}%
Dept. of Mathematics, Kyoto University of Education \\
Fujinomori-cho, Fukakusa, Fushimi-ku\\
Kyoto, 612-8522, Japan\\
TEL: +81-3-1111-1111 \quad FAX: +81-3-1111-1111\\

} \abstract{ Quantum communication is concerned with  the complexity
of entanglement of a state and statistical data analysis is
concerned with the complexity of a model. A common key word for both
is "rank". In this paper we will show that both community is tracing
the same target and that the methods used are slightly different.
Two different methods, the range criterion method from quantum
communication and the determinant polynomial method, are shown as an
examples. }


\maketitle

\section{Introduction}
Quantum communication are strongly interested in entanglement and
classification of entangled states. Multipartite states are
elements, that is, tensors, in some tensor product  space of  the
Hilbert space for each site. On the other hand, in statistical data
analysis, a "tensor" means a multi-array high dimensional data,
which are recently used in various field. Statistical data analysis
are strongly concerned with the data complexity and the model
complexity. They are expressed  by "rank" and "maximal rank"
respectively. Clearly fixing bases in each component Hilbert space,
tensors are expressed its coefficient numerical tensors.
Classification of entanglement under SLOCC is easily seen to be a
classification of the coefficient tensors under mode products.
Classification of  entanglement or determination of maximal rank is
now completed in both community, (in statistical community,  it is
solved from a view point of Jordan form, but somewhat differently in
quantum communication) and the research interest is moved to the
classification of 3 party entanglement, that is, coefficient tensors
with 3 dimensions, $N_{A}, \geq N_{B} \geq N_{C}$. The problem is
known to be notoriously difficult even for  $N_{c}=3.$ for general.
In section 2 we review some terminology in both tensor rank and
quantum information theory. In section 3 we introduce the equivalent
states under SLOCC, as well as some related open problems arising in
quantum information and computation. In section 4 we define a
determinant polynomial(see Sakata-Sumi-Miyazaki(Abstract Book, ISI
2009,Durban)) and show its usefulness as a equivalence criterion. In
section 5 we develop the technique of range criterion and use it to
distinguish inequivalent multipartite states under SLOCC. In section
6 we develop the classification of pure states for the case $3
\times 3 \times 5$, with the introduction of the classification of
the $2 \times m \times n$ case by Lin Chen et all(2006) \cite{CC06}.
For the sake of lack of space,  all are written as possible as
concisely.

\section{Terminologies}

In this section we briefly review basic terminologies of quantum
information theory and tensor rank, for both physicists and
non-physicists. We are especially concerned with the precise
mathematical definition for various physical terminologies.

\subsection{Qubit and quNit}

A quantum state with two levels is called a qubit. A qubit is
mathematically represented as a vector in a 2 dimensional complex
Hilbert space $\bm{H}$. There is a standard basis in $\bm{H}$,
denoted by
\begin{equation}
\bm{e}_{0}=\left(\begin{array}{c}1 \\ 0 \end{array}\right), \ \ \
\bm{e}_{1}=\left(\begin{array}{c}0 \\ 1 \end{array}\right).
\end{equation}
These are denoted by $|0\rangle$ and $|1\rangle$ conventionally.
Therefore, stating precisely, a 1-qubit state is a vector $\bm{x}$
in $\bm{H}$, such that
\begin{equation}
\bm{x}=\alpha |0\rangle+ \beta |1 \rangle
\end{equation}
with the constraint $|\alpha|^{2}+|\beta|^{2}=1$.

Similarly, for a general $N$, a Hilbert space of $N$ dimensional
complex Hilbert space is a state space of $N$-level quantum states,
or quNits. The state is of the form
\begin{equation}
\ket{\ps}=\alpha_{0}|0\rangle +\alpha_{1}|1\rangle + \cdots +
\alpha_{N-1}|N-1\rangle
\end{equation}
with the restriction $|\alpha_{0}|^{2}+\cdots +|\alpha_{N-2}|^{2}+
|\alpha_{N-1}|^{2}=1.$ Here we choose the bases as
$\ket{0}=(1,0,\ldots,0)^T,\ket{1}=(0,1,\ldots,0)^T,\ldots,\ket{n-1}=(0,0,\ldots,1)^T$.
The set of bases is frequently used in quantum information and we
call it computational bases for simplicity. Any state can be
expressed by the computational bases. Of course, we can also choose
different set of bases to decompose the state $\ket{\ps}$. It can be
realized totally by skills in linear algebra.

\subsection{Quantum multipartite states}

The state $\ket{\ps}$ actually describes a local system in physics.
Here the local system can be referred to as one particle, photon,
atom, etc. We can describe a few local systems by using the direct
(Kronecker) product of their states. For example, suppose a few
states $\ket{\ps_i},i=1,2,\ldots,m$ describe the systems $A_i,
i=1,2,\ldots,m$, respectively, then the total system $A_1A_2\ldots
A_m$ is in the state
\begin{equation}
  \label{eq:22} \ket{\Ps}_{A_1A_2\ldots A_m}=\ket{\ps_1}\ox\ket{\ps_2}\ox\ldots\ox\ket{\ps_m}.
\end{equation}
We call it the product multipartite (bipartite for $m=2$) quantum
state. By the language of tensor rank (see next subsection), we say
the state $\ket{\Ps}_{A_1A_2\ldots A_m}$ has tensor rank 1. However,
there are multipartite states which cannot be expressed in Eq.
\ref{eq:2}. For example, the $2\ox2$ bipartite state
\begin{eqnarray}
  \label{eq:bell} \ket{\ps} &=& \ket{0}\ox\ket{0}+\ket{1}\ox\ket{1} \nonumber\\
  & = & \left(\begin{array}{c}
  1\\
  0
  \end{array}\right)\ox\left(\begin{array}{c}
  1\\
  0
  \end{array}\right)+
  \left(\begin{array}{c}
  0\\
  1
  \end{array}\right)\ox\left(\begin{array}{c}
  0\\
  1
\end{array}\right).
\end{eqnarray}
For convenience we will use $\ket{i}\ox\ket{j}=\ket{ij}$ when there
is no confusion. The above state can then be written as
$\ket{\ps}=\ket{00}+\ket{11}$. It's easy to check $\ket{\ps}$ has
tensor rank 2; namely we cannot write that
$\ket{\ps}=\ket{\a}\ox\ket{\b}$. As another example, every trilinear
form in algebraic complexity stands for a tripartite quantum state,
such as the following state
\begin{equation}
  \label{eq:w} \ket{\Ps}_{ABC}=\ket{001}+\ket{010}+\ket{100}.
\end{equation}
This is a typical multipartite state in quantum information and its
tensor rank is 3.

Generally, a multipartite state in the space
$\bm{H}=\bm{H}_{1}\otimes \cdots \otimes  \bm{H}_{m}$ can be written
in terms of the computational bases as follows
\begin{equation}
  \label{eq:multipartite} \ket{\Ps}_{A_1A_2\cdots A_m}=\sum_{i_1,i_2,\cdots,i_m}
  a_{i_1,i_2,\cdots,i_m}\ket{i_1,i_2,\cdots,i_m},
\end{equation}
where the normalization condition keeps\\
$\sum_{i_1,i_2,\cdots,i_m}|a_{i_1,i_2,\cdots,i_m}|^2=1$. A
multipartite state is called an entangled state when it's not of
product state in~\ref{eq:22}, namely

\begin{corollary}
A multipartite pure state is entangled if and only if it has tensor
rank larger than 1.
\end{corollary}

Multipartite states are fundamental resources for a
cross-disciplinary field between quantum physics and information
theory--quantum information and computation theory, which has
developed very fast since 1990s. There have been a rapidly
increasing number of papers contributed to this field because of the
novel ideas and great potential of application and methods to
realize physical and informational tasks. Besides the physics, the
contributors are also from many related areas like mathematics,
chemistry, biology, computer science, engineering and so on. In
recent years, many theoretical plans in quantum information theory
have been realized in experiments. In what follows, we will
introduce the concept of tensor rank and find out its relation to
quantum information.

\subsection{Tensor rank of multipartite states}

Tensor rank is an important concept in data processing, and more
generally in computer science, and has been used in many branches of
science \cite{BCS97}. Expressed in the language of quantum
information, the tensor rank $R(\Psi)$ of a multipartite pure state
$\ket{\Psi}$ in the space $H=H_{A_1}\ox H_{A_2}\ox\cdots\ox H_{A_n}$
is the minimal number $R$ of product states $\ket{\pi_j} =
\bigotimes^{n}_{k=1} \ket{\phi_{jk}}$ with $\ket{\phi_{jk}} \in
H_{A_k}$, $j=1,\ldots,R$, $k=1,\ldots,n$, whose superposition forms
the state $\ket{\Psi}$; that is, $\ket{\Psi} = \sum^{R}_{j=1}
\bigotimes_{k=1}^{n} \ket{\phi_{jk}}$, or equivalently,
\begin{equation}
  \label{eq:tensorrank}
  \ket{\Psi} \in \lin\{ \ket{\pi_1},\ldots,\ket{\pi_R} \}.
\end{equation}

For example, consider two well-known three-qubit states, the
Greenberger-Horne-Zeilinger (GHZ) state and the W state,
\begin{equation}\begin{split}
  \ket{\text{GHZ}} &= \frac{1}{\sqrt2}(\ket{000}+\ket{111}), \\
  \ket{\text{W}}   &= \frac{1}{\sqrt3}(\ket{001}+\ket{010}+\ket{100}).
\end{split}\end{equation}
We need at least two product states to form the GHZ state. So the
tensor rank of $\ket{\text{GHZ}}$ is two. Similarly, one can show
that the tensor rank of $\ket{\text{W}}$ is three. Generally, there
have been techniques for calculating or estimating the tensor rank
of multipartite states in algebraic computational complexity
\cite{AL80}. Note that for bipartite systems, the tensor rank is
identical to the well-known Schmidt rank of a state.

In this case, we can use the tensor rank to characterize a given
quantum state. We will list a few applications and problems in
quantum information by using the results in tensor rank.

\section{Equivalent states under stochastic local
operations and classical communications (SLOCC)}

In quantum physics, physical operators can be denoted as some square
matrix, which may be either non-singular or singular. The general
LOCC operation or a completely positive (CP) map on a state has the
form
\begin{equation}
  \label{eq:locc} \ve(\r)=\sum_iA_i\ox B_i\r A^{\dag}_i\ox
  B^{\dag}_i.
\end{equation}
Notice it actually denotes separable operations, which includes the
set of LOCC operations. The reason is that the standard form of LOCC
is very complicated and not easily handled, so we usually use the
expression in Eq. \ref{eq:locc}. In particular, two states
$\r_1,\r_2$ are one-way equivalent under stochastic LOCC (SLOCC)
when there is some CP map $\ve$ such that $\ve(\r_1)=\r_2$ or
$\ve(\r_2)=\r_1$. If both of them are correct (maybe by virtue of
two CP maps), the two states are equivalent under SLOCC. The reason
we use "stochastic" is that the map may be enforced with a
probability larger than 0 but smaller 1. It is one of the most usual
restrictions in quantum information theory.

As the equivalence problem is quite general and extensive, we only
deal with pure states in this talk. An important problem in quantum
information is to determine whether two pure multipartite states are
(one-way) equivalent or not under SLOCC. Mathematically, two $n$
partite states $\ket{\ps},\ket{\ph}$ are one-way equivalent if and
only if there are some local operators $A_1,\ldots,A_n$ such that
\begin{equation}
  \label{eq:slocc1} \ket{\ps}_{1,\ldots,n}=A_1\ox\ldots\ox
  A_n\ket{\ph}_{1,\ldots,n}.
\end{equation}
Moreover, $\ket{\ps},\ket{\ph}$ are equivalent under SLOCC when the
local operators are non-singular square matrix. That is, we can also
write
\begin{equation}
  \label{eq:slocc} \ket{\ph}_{1,\ldots,n}={A_1}^{-1}\ox\ldots\ox
  {A_n}^{-1}\ket{\ps}_{1,\ldots,n}.
\end{equation}

For example, two states $\ket{00}+\ket{11}$ and $\ket{01}-\ket{10}$
are equivalent because
\begin{equation}
  \label{eq:2} \ket{00}+\ket{11}=\left(\begin{array}{cc}
1&0\\
0&1
\end{array}\right)\ox
\left(\begin{array}{cc}
0&1\\
-1&0
\end{array}\right)(\ket{01}-\ket{10}).
\end{equation}
So the point of deciding two equivalent states is to find out the
direct product of nonsingular matrices that link them. However, this
is usually difficult, even for the tripartite system (c.f. next
paragraph). On the other hand, tensor rank can provide some useful
tool in this context \cite{CDS08}.

\begin{lemma}
  \label{lemma:tr}
  Tensor rank of multipartite states cannot be
  increased under SLOCC.\qed
\end{lemma}

\begin{lemma}
  \label{lemma:tr}
  If $\ket{\ps}$ is one-way equivalent to $\ket{\ph}$, then the tensor
  rank of $\ket{\ps}$ is not smaller than $\ket{\ph}$.\qed
\end{lemma}

\begin{corollary}
  \label{corollary:slocc}
  Two equivalent multipartite states have identical tensor rank.\qed
\end{corollary}

So tensor rank provides a necessary condition for two equivalent
states. For example, the following two tripartite states cannot be
equivalent:
\begin{equation}
  \label{eq:3} \ket{\ph}_{ABC}=\ket{000}+\ket{111}+\ket{222},
\end{equation}
and
\begin{equation}
  \label{eq:4} \ket{\ps}_{ABC}=\ket{012}+\ket{021}+\ket{102}+\ket{120}+\ket{201}+\ket{210}.
\end{equation}
The reason is that the tensor rank of them are 3 and 4 respectively,
see Atkinson-Lloyd's paper in 1979 for details of Eq. \ref{eq:4}. So
the calculation or estimation of two states may imply the
equivalence of them or not. Notice the two states are not one-way
equivalent either. The reason is that operators without full rank
will reduce the local rank of a multipartite state, while the local
rank of both states here are three.

So far we considered the one-copy state. The question will be more
interesting when we study the case of many copies. Given $m$ copies
of $\ket{\ps}$ and $n$ copies of $\ket{\ph}$, we may ask whether its
possible to convert $\ket{\ps}^{\ox m}$ into $\ket{\ph}^{\ox n}$ by
SLOCC. Here, the tensor product state forms a new state in the way
that the corresponding parties are combined together and form a new
party in the new state, i.e.,
\begin{eqnarray}
  \label{eq:tensor}
  \ket{\ps}_{{A_1}\ldots{A_n}}^{\ox m} &=\ket{\ps}_{{A^{1}_1}\ldots{A^1_n}}\ox\ldots\ox
  \ket{\ps}_{{A^m_1}\ldots{A^m_n}}\nonumber\\
  &=\ket{\ps}_{{A^1_1\ldots A^m_1}:{A^1_2\ldots A^m_2}:\ldots:{A^1_n\ldots
  A^m_n}},
\end{eqnarray}
where the new state is still in $n$ parties each of which contains
${A^1_i\ldots A^m_i},i=1,...,n$.

For example, we recall that the W state
$\ket{001}+\ket{010}+\ket{100}$ has tensor rank 3 and the GHZ state
$\ket{000}+\ket{111}$ has tensor rank 2. So it's impossible to make
them equivalent under SLOCC. However, authors have found that we can
realize the one-way equivalence of them when a few copies are
available \cite{CDS08}; e.g., $\ket{GHZ}^{\ox3}$ can be transformed
into $\ket{W}^{\ox2}$ by SLOCC. (The authors deduced that the tensor
rank of two copies of W state is not bigger than 8). It implies we
may realize more equivalence by using of multi-copy states. In a
latest paper \cite{YCG09}, authors have proved that the tensor rank
of two-copies of W states analytically equals 7, which is a
surprising fact. It's going to be an interesting problem to address
the tensor rank of $\ket{W}^{\ox n}$.

\section{A necessary condition by determinant
polynomials of SLOCC-equivalence of tensors with 3-slices}

In this section we introduce a necessary condition of deciding the
equivalent multipartite states under SLOCC, by using the so-called
3-slices' determinant polynomials developed by
Sakata-Sumi-Miyazaki(Abstract Book, ISI 2009,Durban).

Let matrices $A,B$ and $C$ be $n \times n$ matrices,  and we call a
triple $(A,B,C)$ as a $n \times n \times 3$ tensor and denote by
$T=A:B:C$. Then each matrices $A$, $B$ and $C$ is called an $i$-th
slice of the tensor $T$, $i=1,2,3,$ respectively. From a data
analytic point of view, for a tensor $T$, we are mainly concerned
with its tensor rank, denoted by $rank(T)$ which is the smallest
number of rank 1 tensors by which the tensor is expressed as a sum
(notice it coincides with the concept in Section 2.3). Therefor
tensor rank is an index of complexity of data.

For a tensor $T=A:B:C$ we consider the two types of transformation,
\begin{itemize}
\item[Type 1]  $(A,B,C) \rightarrow (PAQ,PBQ,PCQ)$ with nonsingular matrices $P$ and $Q$
\item[Type 2]  $(A,B,C) \rightarrow  (g_{11}A+g_{21}B+g_{31}C,g_{12}A+g_{22}B+g_{32}C,g_{13}A+g_{32}B+g_{33}C),$
where $G=(g_{ij})$ is a nonsingular matrix.
\end{itemize}
\begin{definition}
\label{de:equivalenttensor} Two tensor $T_{1}$ and $T_{2}$  are said
to be equivalent if $T_{1}$ and $T_{2}$ are inter-convertible by a
sequence of transformations of type 1 and type 2.
\end{definition}
\begin{remark}
Tensor rank is invariant under both type of transformations.
\end{remark}

Subsequently, we are going to find out the connection between a $n
\ox n \ox 3$ state and its corresponding 3-slices tensor. Let us
consider two $n \ox n \ox 3$ states
\begin{equation}
\ket{\Ps}=\sum^2_{k=0} \left(\sum^{n-1}_{i,j=0}a_{ijk}\ket{i,j}
\right)\ket{k}
\end{equation}
and
\begin{equation}
\ket{\Ph}=\sum^2_{k=0} \left(\sum^{n-1}_{i,j=0}b_{ijk}\ket{i,j}
\right)\ket{k}.
\end{equation}
According to the definition of SLOCC equivalence in
Eq.~\ref{eq:slocc1}, there should be three nonsingular matrices $A,
B$, $C$ such that $\ket{\Ps}=A \ox B \ox C\ket{\Ph}$, namely
\begin{eqnarray}
\label{ar:sloccequivalent} A \ox B
\left(\sum^{n-1}_{i,j=0}a_{ijk}\ket{i,j} \right) &=& \sum^2_{l=0}
g_{lk} \left(\sum^{n-1}_{i,j=0}b_{ijl}\ket{i,j}
\right),\nonumber\\
k &=& 0,1,2,
\end{eqnarray}
where $G=[g_{ij}]^{3\times3}$ is nonsingular. If we perform the
partial rotation on system B by column row $\ket{i}\rightarrow
\bra{i}$ and define two tensors
\begin{eqnarray}
\label{ar:tensor1}
  T_1=\sum^{n-1}_{i,j=0}a_{ij0}\ket{i}\bra{j}:
  \sum^{n-1}_{i,j=0}a_{ij1}\ket{i}\bra{j}:
  \sum^{n-1}_{i,j=0}a_{ij2}\ket{i}\bra{j},
\end{eqnarray}
and
\begin{eqnarray}
\label{ar:tensor2}
  T_2=\sum^{n-1}_{i,j=0}b_{ij0}\ket{i}\bra{j}:
  \sum^{n-1}_{i,j=0}b_{ij1}\ket{i}\bra{j}:
  \sum^{n-1}_{i,j=0}b_{ij2}\ket{i}\bra{j}.
\end{eqnarray}
Then Eq.~\ref{ar:sloccequivalent} states that tensors $T_1$ and
$T_2$ are equivalent in terms of
definition~\ref{de:equivalenttensor}. In this sense, the equivalence
of two tensors will immediately cause the equivalence of two
tripartite states under SLOCC.

However, it should be noted that tensor rank does not determine
equivalent class, and that it is merely an invariant under SLOCC. If
$T_{1}$ and $T_{2}$ are equivalent under SLOCC, the rank of them are
the same. Thus the rank is used as a test function of equivalence,
that is, if the ranks are not identical, the two states are not
equivalent. So, tensor rank is a necessary condition for
equivalence. However, it is too coarse as an index of equivalence
classes. In fact, the number of tensor rank is extremely less than
the number of equivalent classes under SLOCC in general.

In this paper we propose a new index of equivalence under SLOCC,
which is more fine than tensor rank as an index of equivalence. The
index is the polynomial determinant of a tensor. For tensor
$T=A:B:C$ we define the polynomial by
\begin{equation}
f(x,y,z)=det(xA+yB+zC),
\end{equation}
which we call the determinant polynomial of the tensor $T$. By the
results in Sakata-Sumi-Miyazaki(Abstract Book, ISI 2009,Durban), we
have
\begin{theorem}\label{thm1.10}
Two tensors $T_{1}$ and $T_{2}$ are equivalent only if, based on the
two monic polynomials obtained from both determinant polynomials,
the algebraic equation about a nonsingular matrix $G$ defined by
\begin{equation}
f_{2}(\bm{x})=f_{1}(\bm{x}G^{t}),
\end{equation}
has at least one solution $G$. \qed
\end{theorem}

\begin{example}
(Sakata-Sumi-Miyazaki(Abstract Book, ISI 2009,Durban)) We use
theorem~\ref{thm1.10} to derive the equivalence of two $3 \times
3\times 3$ tensors
\begin{equation}
T_{1}=\left( \begin{array}{ccc}
  1 & 0 & 0  \\
   0 & 0 & 0 \\
 0 & 0 & 0 \\
\end{array}
\right); \left( \begin{array}{cccc}
  0 & 0 & 0  \\
  0 & 1 & 0 \\
  0 & 0 & 0 \\
\end{array}
\right); \left( \begin{array}{cccc}
   0 & 0 & 0  \\
   0 & 0 & 0 \\
 0 & 0 & 1 \\
\end{array}
\right),
\end{equation} and
\begin{equation}
T_{2}=\left( \begin{array}{ccc}
  1 & 0 & 0  \\
   0 & 1 & 0 \\
 0 & 0 & 0 \\
\end{array}
\right); \left( \begin{array}{cccc}
  0 & 0 & 0  \\
  0 & 1 & 0 \\
  0 & 0 & 1 \\
\end{array}
\right); \left( \begin{array}{cccc}
   1 & 0 & 0  \\
   0 & 0 & 0 \\
 0 & 0 & 1 \\
\end{array}
\right).
\end{equation}

Let $f_{1}$ and $f_{2}$ be the determinant polynomials of $T_{1}$
and $T_{2}$ respectively. Then we have
\begin{equation}
f_{1}(x,y,z)=xyz,
\end{equation}
and
\begin{equation}
f_{2}(x,y,z)=(x+y)(y+z)(z+x).
\end{equation}
So, by the transformation
\begin{equation}
x \rightarrow (x-y+z)/2, \ \  y \rightarrow (x+y-z)/2,  \ \ z
\rightarrow (-x+y+z)/2,
\end{equation}
we have $f_{2} \rightarrow f_{1}$. So the equivalence of $T_{1}$ and
$T_{2}$ is not denied. Carefully looking, we see that they are
equivalent. In fact, after subtracting $B_{2}$ from  $A_{2}$ and
adding $C_{2},$ dividing $A_{2}$ by $2$, $A_{2}$ becomes $A_{1}$.
Then by subtracting $A_{2}$ from $C_{2}$, $C_{2}$ becomes $C_{1}$.
Finally by subtracting $C_{2}$ from $B_{2}$, $B_{2}$ becomes
$B_{1}.$ Thus, $T_{1}$ and $T_{2}$ are equivalent.
\end{example}

\section{Classification of SLOCC equivalent \\ states based on range criterion}

Classification of multipartite entangled states is an important
topic in quantum information. There have been many papers in this
topic over past years \cite{HHH09}. Especially the classification of
K partite $(K\geq3)$ entangled states under SLOCC is known to be
very difficult. First we introduce the necessary and sufficient
condition for equivalence under SLOCC which appeared in the paper of
Lin Chen et al(2006) \cite{CC06}. To explain our method, we need to
build more basic concepts in quantum information.

\subsection{Density matrix and reduced density matrix of a state}

\begin{definition}
Let $\bm{H}$ be $N_{1}$dimensional complex Hilbert space and
$|\bm{x} \rangle$ be an element of it, that is, a state. Then
\begin{equation}
\rho=|\bm{x}\rangle \langle \bm{x}|
\end{equation}
is called as the density
matrix of the state $|\bm{x}\rangle$.
\end{definition}

\begin{proposition}
\begin{equation}
Tr(\rho)=||\bm{x}||^{2}=1.
\end{equation}
\end{proposition}

\begin{definition}
Assume that $|\bm{x}_{1}\rangle, \cdots, |\bm{x}_{K}\rangle$ are a
collection of $K$ pure states and
$p_{1},...,p_{K}(\sum_{i=1}^{K}p_{i}=1)$ is a probability. Then the
mixed state density matrix is defined as
\begin{equation}
\rho=\sum_{i=1}^{K}p_{i}\rho_{\bm{x}_{i}}.
\end{equation}
\end{definition}
Properties of density matrices are listed below.
\begin{itemize}
\item[1]  A pure sate is defined as both a vector and a density matrix, however, a mixed sate is defined only through a density matrix, that is, through a set of pure states and probabilities.
\item[2]  A quantum state always corresponds to a density operator one-to-one. When two pure states are proportional, e.g., $|0\rangle$ and $e^{i \alpha}|0\rangle$, they stand for the same state, for the global phase $e^{i \alpha}$ does not lead to difference in physics.
For the case of mixed state which is always a matrix, there will not
be global phase otherwise it's not positive semidefinite. For
instance, if $\rho$ is a density matrix, $e^{i \alpha}\rho$ is not
legal by the definition of density operators.
\end{itemize}

\begin{definition}
Partial trace is a method of obtaining the reduced density matrix,
that is, the  marginal density matrix, of Alice and Bob
respectively. Alice's reduced density matrix and Bob's reduced
density matrix are obtained respectively by
\begin{equation}
Tr_{B}^{A}|\bm{\Phi} \rangle \langle \bm{\Phi}|=\sum_{j}(\bm{I}
\otimes \langle \bm{j}|) |\bm{\Phi} \rangle \langle \bm{\Phi} |
((\bm{I} \otimes | \bm{j}\rangle)
\end{equation} and
\begin{equation}
Tr_{A}^{B}|\Phi \rangle \langle \bm{\Phi}|=\sum_{i}( \langle \bm{i}|
\otimes \ \bm{I})  |\bm{\Phi} \rangle \langle \bm{\Phi} |( \langle
\bm{i}| \otimes \bm{I}).
\end{equation}
where $\{|\bm{i}\rangle \}$ and $\{|\bm{j}\rangle \}$ are
orthonormal bases of $H_{A}$ and $H_{B} respctively.$
\end{definition}
\begin{remark}
Partial trace is also extendedly defined for joint mixture states by
linearity. We also denote reduced density matrix as reduced density
in this paper.
\end{remark}

\begin{definition}
Let $|\Phi\rangle \in \bm{H}_{A}\otimes \bm{H}_{B}\otimes
\bm{H}_{C}$ be the joint state of Alice, Bob and Cherry. Then the
density matrix of the joint state is  $|\Phi\rangle \langle \Phi |$
and  Alice's reduced density is defined by
\begin{equation}
Tr_{BC}^{C}|\bm{\Phi} \rangle \langle
\bm{\Phi}|=\sum_{j}\sum_{k}(\bm{I} \otimes \langle \bm{j}| \otimes
\langle \bm{k}|) |\bm{\Phi} \rangle \langle \bm{\Phi} | ((\bm{I}
\otimes |\bm{j}\rangle \otimes |\bm{k} \rangle)
\end{equation}
where $\{|\bm{j}\rangle \}$ and  $\{|\bm{k}\rangle \}$ are
orthonormal bases of $H_{B}$ and $H_{C}.$
\end{definition}
\begin{remark}
$Tr_{AC}^{B}$,$Tr_{AB}^{C}$, $Tr_{A}^{BC}$,$Tr_{B}^{AC}$, and
$Tr_{C}^{AB}$ is also defined similarly.
\end{remark}

\subsection{local rank}

The local rank of a multipartite state means the ranks of reduced
density operators; For example, the state
$|\bm{000}\rangle+|\bm{111}\rangle$ has local rank $2,2,2;$ and the
state
$|\bm{000}\rangle+(|\bm{1}\rangle+|\bm{2}\rangle)(|\bm{3}\rangle+|\bm{4}\rangle+|\bm{7}\rangle)(|\bm{5}\rangle+|6\rangle-|0\rangle)$
still has local ranks 2,2,2. In fact the latter has the reduced
density operator $|\bm{0}\rangle\langle
\bm{0}|+a(|\bm{1}\rangle+|\bm{2}\rangle)(\langle \bm{1}|+\langle
\bm{2}|)$, $|\bm{0}\rangle\langle
\bm{0}|+b(|\bm{3}\rangle+|\bm{4}\rangle+|\bm{7}\rangle)(\langle
\bm{3}|+\langle \bm{4}|+\langle \bm{7}|)$, and $|\bm{0}
\rangle\langle
\bm{0}|+c(|\bm{5}\rangle+|\bm{6}\rangle-|\bm{0}\rangle)(\langle
\bm{5}|+\langle \bm{6}|-\langle \bm{0}|)$, for each system
respectively, where a,b,c are non-zero constants that can be
calculated. Evidently, they all have rank 2; and this is the meaning
of local ranks.

\subsection{Range criterion(Lin Chen et al, \cite{CC06})}

\begin{definition}
Let $\rho=|\bm{\Psi} \rangle \langle \bm{\Psi}|$ be a joint density
given by a pure state $\bm{\Psi} \rangle \in \bm{H}_{A} \otimes
\bm{H}_{B} \otimes \bm{H}_{C}$. Then, for clarification, the
marginal density for each party are denoted by
$\rho^{A}_{\Psi_{ABC}}$,  $\rho^{B}_{\Psi_{ABC}}$,
$\rho^{C}_{\Psi_{ABC}}$,respectively.
\end{definition}

\begin{theorem} (Range criterion). \label{rangecriterion}
Let $\bm{\Psi}_{ABC}$ and $\bm{\Phi}_{ABC}$ are two pure states in
$\bm{H}_{A} \otimes \bm{H}_{B} \otimes \bm{H}_{C}$ such that
$\bm{\Psi}_{ABC}=V_{A} \otimes V_{B}\otimes V_{C} \bm{\Phi}_{ABC}$.
Let $S_{1}=\{|\bm{x} \rangle_{BC}\in
R(\rho_{\bm{\Psi}_{ABC}}^{BC})\}$ and $S_{2}=\{|\bm{x} \rangle_{BC}
\in V_{B}\otimes V_{C}R(\rho_{\Phi_{ABC}}^{BC})\}$ Then
\begin{enumerate}
\item all local ranks are equal
\item  $S_{1}=S_{2}$
\end{enumerate}
This is a necessary and sufficient condition for equivalence of both
states.
\end{theorem}
\begin{corollary}
If two pure states of a multipartite system are equivalent under
SLOCC, the number of linearly independent product states in the
range of the adjoint reduced density matrices of each party of them
must be equal.
\end{corollary}

\begin{example}
Consider the well-known GHZ and W state. One can check that there
are two product states $\ket{00},\ket{11}$ in $R(\rho_{GHZ}^{BC})$,
while there is only one $\ket{00}$ in $R(\rho_{W}^{BC})$. So we
conclude that GHZ and W states are not equivalent under SLOCC.  \qed
\end{example}

\section{Classification of $2\times M\times N$ states and application to $3\times 3\times 5$ states}

By using the range criterion in theorem~\ref{rangecriterion}, we
already developed a technique to classify inequivalent states of
$2\times M\times N$ system as follows.
\begin{theorem} (Lin-Chen et al \cite{CC06})
\label{thm:2mn} we have\\
$\left|\Psi\right\rangle_{2\times M \times
N}\sim \left\{\begin{array}{l}
\left|\Omega_0\right\rangle\equiv(a\left|0\right\rangle
+b\left|1\right\rangle)\left|M-1,N-1\right\rangle
\\+\left|\Psi\right\rangle_{2\times(M-1) \times(N-1)},\\
\left|\Omega_1\right\rangle\equiv\left|0,M-1,N-1\right\rangle
\\+\left|1,M-1,N-2\right\rangle+\left|\Psi\right\rangle_{2\times(M-1)\times(N-2)},\\
\left|\Omega_2\right\rangle\equiv\left|\Omega_0\right\rangle
+\left|0,M-1\right\rangle\left|\chi\right\rangle,b\neq0,\\
\left|\Omega_3\right\rangle\equiv\left|\Omega_0\right\rangle
+\left|1,M-1\right\rangle\left|\chi\right\rangle,a\neq0.
\end{array}\right.$\\
Here $\ket{\chi}=\sum^{N-2}_{i=0} a_i\ket{i}$ as a random state. The
condition $a\neq0$ or $b\neq0$ keeps $\left|\Omega_2\right\rangle$
and $\left|\Omega_3\right\rangle$ not becoming
$\left|\Omega_0\right\rangle$.
\end{theorem}
Such equivalence relation shows that the lower rank classes of the
entangled states can be used to generate the higher rank classes of
the true entangled states for any $2\times M\times N$ system, called
as "Low-to-High Rank Generating Mode" or LHRGM for short. So the
corollary and the range criterion of the theorem 1 provide a
systematic method to classify all kinds of true tripartite entangled
states in the $2\times M\times N$ system under SLOCC in \cite{CC06}.

In particular, we can use the classification of $2\times M \times N$
states with $M\leq3, N\leq6$ in \cite{CC06} to derive the tensor
rank of $3\times 3 \times 5$ states, which is either 6 or 7 by
Atkinson et al \cite{AL80}. By using a tensor product of three
nonsingular matrices on the system, every $3\ox3\ox5$ state can be
written as
\begin{equation}
  \label{eq:1}
  \ket{\Ps}_{ABC}=\ket{\ps}+\ket{2}(\ket{0}\ket{\a}+\ket{1}\ket{\b}+\ket{2}\ket{\g}),
\end{equation}
where $\ket{\ps}$ is a $2\ox n\ox p$ state with $n\leq3,2\leq
p\leq5$. So it's possible to infer the tensor rank of
$\ket{\Ps}_{ABC}$ by simplifying the expression in Eq. \ref{eq:1},
while which is already classified in \cite{CC06}. For convenience,
we list the inequivalent states derived thereof, where readers can
check the details. At present we are still studying this problem and
new results will be reported later.

\begin{tabular}{l|l}
systems' ranks & Inequivalent states under SLOCC\\
\hline $2\times3\times 6$ &
$\left|000\right\rangle+\left|011\right\rangle+\left|022\right\rangle+\left|103\right\rangle$\\
&$+\left|114\right\rangle+\left|125\right\rangle;$ \\
\hline$2\times3\times 5$ &
$\left|024\right\rangle+\left|000\right\rangle+\left|011\right\rangle$\\
&$+\left|102\right\rangle+\left|113\right\rangle;$\\
&$\left|024\right\rangle+\left|121\right\rangle+\left|000\right\rangle+\left|011\right\rangle$\\
&$+\left|102\right\rangle+\left|113\right\rangle;$\\
\hline$2\times3\times 4$ &
$\left|123\right\rangle+\left|012\right\rangle+\left|000\right\rangle+\left|101\right\rangle;$\\
&$\left|023\right\rangle+\left|012\right\rangle+\left|000\right\rangle+\left|101\right\rangle;$\\
&$\left|123\right\rangle+\left|012\right\rangle+\left|110\right\rangle
+\left|000\right\rangle+\left|101\right\rangle;$\\
& $\left|023\right\rangle+\left|122\right\rangle
+\left|012\right\rangle+\left|000\right\rangle+\left|101\right\rangle;$\\
& $\left|023\right\rangle+\left|122\right\rangle
+\left|012\right\rangle+\left|110\right\rangle$ \\
& $+\left|000\right\rangle+\left|101\right\rangle;$\\
\hline$2\times3\times 3$ &
$\left|000\right\rangle+\left|111\right\rangle+
\left|022\right\rangle;$\\
& $\left|000\right\rangle+\left|111\right\rangle+
\left|022\right\rangle+\left|122\right\rangle;$\\
& $\left|010\right\rangle+\left|001\right\rangle+
\left|112\right\rangle+\left|121\right\rangle;$\\
&$\left|100\right\rangle+\left|010\right\rangle+\left|001\right\rangle+
\left|112\right\rangle+\left|121\right\rangle;$\\
& $\left|100\right\rangle+\left|010\right\rangle+
\left|001\right\rangle+\left|022\right\rangle;$\\
& $\left|100\right\rangle+\left|010\right\rangle+
\left|001\right\rangle+\left|122\right\rangle;$\\
\hline$2\times3\times 2$ &
$\left|000\right\rangle+\left|011\right\rangle+\left|121\right\rangle;$\\
\end{tabular}

\begin{tabular}{l|l}
&$\left|000\right\rangle+\left|011\right\rangle+\left|110\right\rangle+\left|121\right\rangle;$\\
\hline $2\times2\times 4$ &
$\left|000\right\rangle+\left|011\right\rangle+\left|102\right\rangle+\left|113\right\rangle;$\\
\hline $2\times2\times 3$ &
$\left|000\right\rangle+\left|011\right\rangle+\left|112\right\rangle;$\\
&$\left|000\right\rangle+\left|011\right\rangle+\left|101\right\rangle+\left|112\right\rangle;$\\
\hline $2\times2\times 2$ &
$\left|000\right\rangle+\left|111\right\rangle;$\\
&$\left|001\right\rangle+\left|010\right\rangle+\left|100\right\rangle;$\\
\hline $1\times3\times 3$ &
$\left|000\right\rangle+\left|011\right\rangle+\left|022\right\rangle;$\\
\hline $1\times2\times 2$ &
$\left|000\right\rangle+\left|011\right\rangle;$\\
\hline $2\times1\times 2$ &
$\left|000\right\rangle+\left|101\right\rangle;$\\
\hline $2\times2\times 1$ &
$\left|000\right\rangle+\left|110\right\rangle;$\\
\hline $1\times1\times 1$ & $\left|000\right\rangle.$
\end{tabular}

\section{Conclusions}

In this paper we investigated a few problems between tensor rank and
quantum information theory. We proposed the 3-sliced tensor to
address the SLOCC-equivalence problem. We also have shown that the
range criterion can help find out inequivalent states and thus
tensors, which is likely to help compute the tensor rank of
$3\times3\times5$ tensors.

The open problems could be that how to find out the sufficient
condition by using 3-sliced tensors. Besides, it's also interesting
to compute the multi-copies of W states, which is an important
resource in quantum information, in both theory and experiment.
Nevertheless, we only said some marginal applications of tensor rank
to quantum information due to the restriction of space, as we
already proposed many other connections arising from quantum
information (e.g., entanglement measure \cite{EB06}, separability
problem, construction of positive partial transpose entangled states
and so on). We will propose more results based on the connections
between these two rapidly developing fields.

The Center for Quantum Technologies is funded by the Singapore
Ministry of Education and the National Research Foundation as part
of the Research Centres of Excellence programme. Toshio Sakata,
Toshio Sumi, and Mitsuhiro Miyazaki are financially supported by
Japanese Science Promotion Project(B) No. 20340021.

\end{document}